\PassOptionsToPackage{table}{xcolor}
\documentclass[manuscript,screen,nonacm]{acmart}

\usepackage{caption}

\usepackage{graphicx}
\usepackage{float} 
\usepackage{subfig}
\usepackage{subcaption}
\usepackage{overpic}

\usepackage{booktabs}
\AtBeginDocument{%
  }

\begin{document}


\title{Scent for Vent: Exploring Traditional Chinese Medicine Scents on Driving Fatigue in Autonomous Vehicles}


\author{Liang Luo}
\authornote{These authors contributed equally to this work.}
\affiliation{%
  \department{School of Computer Science and Engineering}
  \institution{University of Electronic Science and Technology of China}
  \city{Chengdu}
  \country{China}
}
\email{luoliang@uestc.edu.cn}

\author{Nengyue Su}
\authornotemark[1]
\affiliation{%
  \department{School of Computer Science and Engineering}
  \institution{University of Electronic Science and Technology of China}
  \city{Chengdu}
  \country{China}
}

\author{Yu Gu}
\affiliation{%
  \department{School of Computer Science and Engineering}
  \institution{University of Electronic Science and Technology of China}
  \city{Chengdu}
  \country{China}
}
\email{yugu.bruce@ieee.org}

\author{Jiuyi Wang}
\affiliation{%
  \department{School of Computer Science and Engineering}
  \institution{University of Electronic Science and Technology of China}
  \city{Chengdu}
  \country{China}
}

\author{Fuji Ren}
\affiliation{%
  \department{School of Computer Science and Engineering}
  \institution{University of Electronic Science and Technology of China}
  \city{Chengdu}
  \country{China}
}
\email{renfuji@uestc.edu.cn}

\begin{abstract}
Driving fatigue poses a critical threat to driving safety, especially in autonomous driving. While scent modulation alleviates it effectively, Traditional Chinese Medicine (TCM) scents---which are associated with emotions---have not yet been studied in this context. This study explores TCM scents’ impact on driving fatigue via two experiments: Study 1 maps emotion-related TCM scents onto an arousal-valence space, empirically confirming their emotional influence; Study 2 uses a simulator to evaluate Argy Wormwood and Tangerine Peel across alertness, driving performance, and acceptance. Results show Argy Wormwood, with potent alertness but low comfort, suits short-term warnings; Tangerine Peel, moderately alerting but comfortable, fits long-term use. To the best of our knowledge, we are the first to quantify the TCM scents’ emotional impact and their application in driving fatigue mitigation, these findings offer a novel, non-intrusive approach to designing in-car interactive environments aimed at improving driver well-being and road safety.
\end{abstract}



\begin{teaserfigure}
  \centering
  \includegraphics[width=\textwidth]{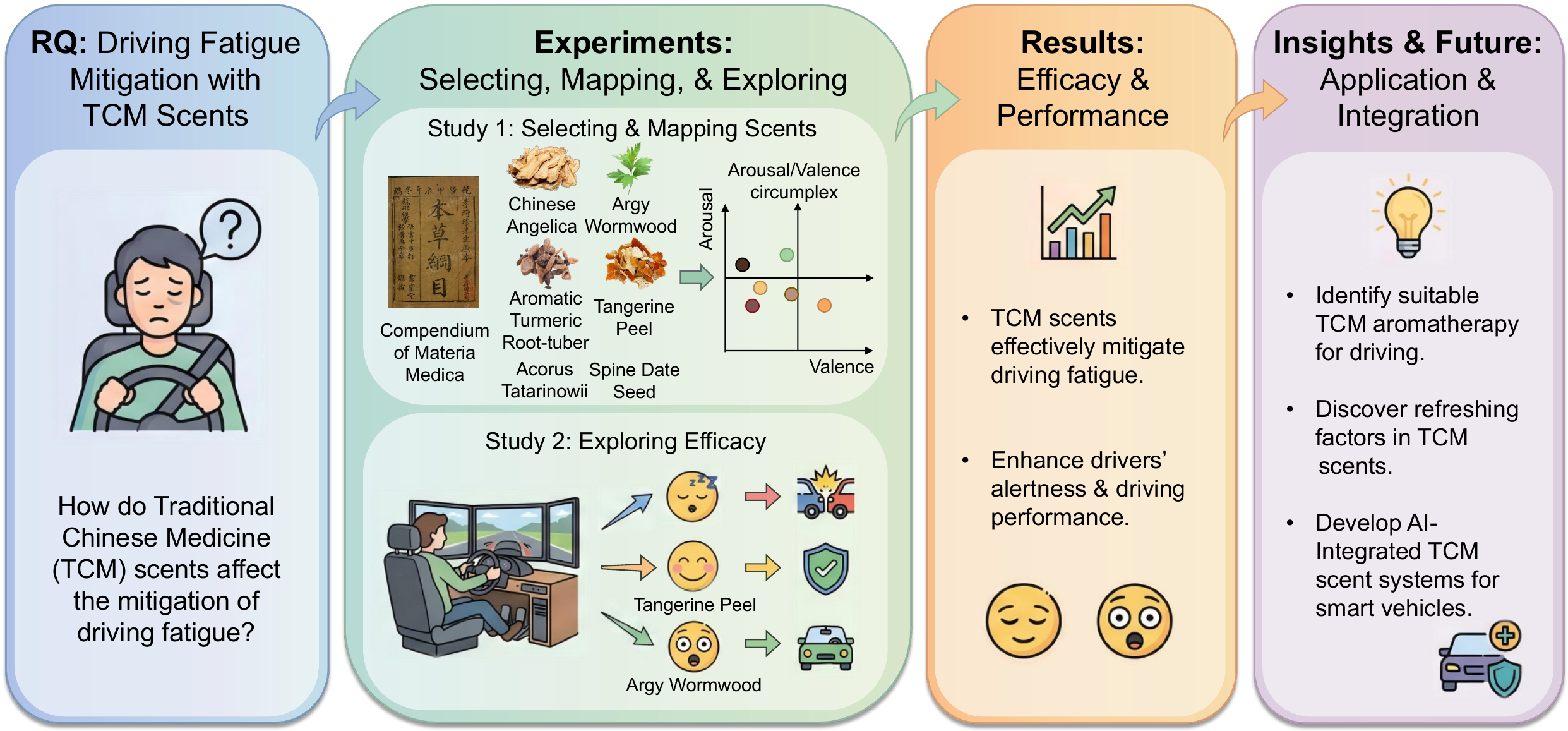}
  \caption{
    Our research demonstrates that Traditional Chinese Medicine scents, such as Tangerine Peel and Argy Wormwood, can effectively mitigate driving fatigue---a critical issue in autonomous driving that impacts both safety and comfort.
    Specifically, Tangerine Peel enhances alertness and comfort, whereas Argy Wormwood boosts alertness despite slightly lower comfort ratings.
    The illustration was generated using Google Gemini with a prompt depicting a drowsy driver in an autonomous vehicle and a simulated driving cockpit.
  }
  \label{fig:1}
\end{teaserfigure}

\maketitle

\section{Introduction}
Currently, the emergence of autonomous driving technology has made driving more monotonous, which in turn increases the likelihood of drivers experiencing passive fatigue (as shown in Figure~\ref{fig:1}) \cite{jamson2013behavioural, ZWAHLEN2016183, HALLVIG201418, phillips2015review}. This mental state, characterized by low valence and low arousal, prolongs drivers' reaction times \cite{eyben2010emotion}. If left unaddressed, it can even trigger traffic accidents \cite{li2024review}, posing significant risks to road safety \cite{zhang2016traffic}.

Sensory stimulation is a common approach to mitigating driving fatigue, and such stimulation can be categorized into visual \cite{li2021visual,qin2021characteristics,hassib2019detecting}, auditory \cite{orsini2024music,trumbo2017name,schwarz2012car}, and olfactory \cite{dahlman2024vehicle,jiang2023study,okazaki2018safe} techniques. Given the pivotal roles of vision and hearing in driving tasks, olfactory modulation offers unique advantages in preserving drivers' functional \cite{PMID:24830042}. Existing scent-related research has predominantly focused on Western essential oils, which may lack broad cultural resonance among Chinese populations. However, insights from traditional Chinese aromatherapy practices \cite{nan2013aromatherapy,giannenas2020history} suggest that scents derived from Traditional Chinese Medicine (TCM) could be more culturally acceptable. Despite this potential, the impact of TCM scents on human emotions and driving fatigue remains underexplored. This study therefore aims to investigate this understudied domain.

More specifically, two studies involving Chinese participants were conducted using a high-fidelity driving simulator. In Study 1, six TCM scents were mapped to the arousal/valence circumplex (as in \cite{dmitrenko2020caroma}). Among these, Argy Wormwood (with the highest arousal level) and Tangerine Peel (with the highest valence level) were selected for further investigation. Study 2 aimed to explore the specific effects of these two scents on driving fatigue using a pre-designed autonomous driving scenario: prolonged driving on a straight highway with no traffic under cloudy weather conditions. The results show that both scents can improve driver alertness and reduce reaction times. However, their applications differ: Argy Wormwood is suitable for short-term use due to its high intensity yet low acceptance, while Tangerine Peel is preferred for long-term application owing to its higher popularity.  

This research examined the effect of Traditional Chinese Medicine (TCM) scents on mitigating driving fatigue, yet several questions still remain unresolved. These include the limited duration of scent effectiveness in aromatherapy, as well as the comparative performance between TCM scents and conventional floral or fruity scents. Furthermore, the question of how driving fatigue can be detected promptly remains unanswered. Additionally, it is unclear whether it is feasible to develop a scent-based fatigue mitigation system that predicts the onset of driving fatigue by analyzing driving contexts (such as the scenario adopted in our driving simulator) via artificial intelligence (AI).

\textbf{Contributions:}  
(1) Screened out Traditional Chinese Medicine (TCM) scents associated with emotion regulation, and mapped these scents onto the arousal-valence model through experimental methods. (2) Conducted driving simulator experiments to investigate the influence of TCM scents on driving across three dimensions: driver alertness, driving performance, and scent acceptance. (3) Deduced the differentiated application scenarios of different TCM scents based on their respective characteristics, and validated these inferences through participant interviews.

\textbf{Novelty:}  
(1) To our knowledge, this study is the first to quantify the emotional effects of Traditional Chinese Medicine (TCM) scents by mapping them into the arousal-valence dimensional space. Previous studies have mainly focused on experimentally verifying the efficacy of TCM in treating emotional disorders. (2) This study is the first to explore the role of TCM scents in driving fatigue scenarios, clarifying that scents with different arousal and valence levels are suitable for different application scenarios. By contrast, existing studies have mostly adopted common Western single or blended essential oils as research objects.

\textbf{Significance:} 
Based on existing technologies, this study proposes a feasible solution to the driving fatigue problem caused by autonomous driving, which opens up a new direction for regulating driving fatigue through scent intervention. For young drivers with Chinese cultural backgrounds, Traditional Chinese Medicine (TCM) scents can effectively alleviate driving fatigue, and thus help ensure driving safety.



\section{Related Work}
We structured the related work around two research areas:

\subsection{How to mitigate driving fatigue?}
Based on the classification of sensory stimuli, methods aimed at mitigating driving fatigue can be categorized into visual, auditory, and olfactory approaches. With regard to visual methods, adjustments to ambient lighting have been employed to enhance driver alertness \cite{li2021visual,qin2021characteristics,hassib2019detecting}. For auditory methods, music \cite{guo2024could,orsini2024music} or name music game \cite{trumbo2017name} have proven effective. In driving scenarios, vision and hearing are the primary channels for drivers to receive information. Although modulation through these two sensory modalities can cause driver distraction, olfaction is far less frequently used during driving and has a closer connection with emotional states \cite{alaoui1997basic}. For this reason, this study uses olfactory modulation as an approach to alleviate driving fatigue.

Current research has confirmed that scents including mint, grapefruit, lemon, and lavender can significantly improve driving fatigue \cite{jiang2024scented,dahlman2024vehicle,yoshida2011study}. However, existing studies have mostly focused on common flower and fruit scents that are widely applied in Western countries. Due to differences in cultural backgrounds and physiological characteristics, these scents are not commonly used by the Chinese population \cite{yunjun2013perfume}. Therefore, it is urgent to investigate scents that are more familiar and acceptable to Chinese people.

\subsection{Why we choose TCM?}
The Chinese people have long held the traditional practice of using Traditional Chinese Medicine (TCM) to regulate emotions since ancient times \cite{farrar2020clinical}. Within the theoretical framework of TCM, there is a bidirectional causal association between the smooth flow of \textit{qi}\footnote[1]{Chinese identify \textit{qi} as the natural energy intrinsic to all things that exist in the universe, as the fundamental life energy responsible for health and vitality\cite{mccaffrey2003qigong}.} and both healthy mental status and emotional balance \cite{zhou2021conceptualization}. Specific Chinese medicinal herbs, including argy wormwood \cite{khan2016gaba, melo2011antidepressant}, Chinese angelica \cite{li2016effects}, acorus tatarinowii \cite{rai2023role, ilaiyaraja2012effect}, spine date seed, and tangerine peel \cite{mo2017effects}, have been proven effective in the treatment of depression \cite{li2020traditional,zhang2019challenge,sun2022dissecting, zhou2016new}. However, most TCM preparations are intended for internal use. Relatively few TCM formulas have been processed into high-purity essential oils for aromatherapy applications, and the specific effects of TCM aromatic scents on emotional states remain unclear. For this reason, this study selects TCM essential oils with potential links to emotion regulation for further investigation.

\section{Experiments}

\subsection{Overview of Experiments}
To explore the effects of TCM scents on mitigating driving fatigue, we designed two studies.
\begin{itemize}
    \item \textbf{Study 1:} Mapping TCM scents onto the arousal/valence circumplex. 
    \item \textbf{Study 2:} Exploring the effects of TCM scents on driving fatigue.
\end{itemize}
Both studies have obtained ethical approval from the institution's ethics committee\protect\footnotemark[2]. All participants were carefully screened to make sure they had no breathing problems and no smell allergies. The selection method and process of participants \cite{koch2023leveraging, deuber2025moving} are detailed in Appendix A.

\footnotetext[2]{Due to the requirement of anonymity for manuscript submission, the name of the institution is temporarily not disclosed.}

\subsection{Study 1: Mapping TCM scents onto the arousal/valence circumplex.}

\begin{figure}
    \begin{minipage}[]{0.48\textwidth}
        \vspace{0.0in}
        \centering
        \resizebox{1\linewidth}{!}{
            \begin{tabular}{|l|l|l|}
            \hline
            TCM                                                                                                                    & Ingredients                                                             & Efficacy                                                                                                                              \\ \hline
            \cellcolor[HTML]{F2F2F2}{\color[HTML]{08090C} Argy Wormwood}                                                           & \begin{tabular}[c]{@{}l@{}}Eucalyptol, \\ Caryophyllene\end{tabular}    & \begin{tabular}[c]{@{}l@{}}Warm the meridians to stop bleeding, \\ dispel cold and relieve pain, \\ calm anxiety\end{tabular}         \\ \hline
            \cellcolor[HTML]{F2F2F2}{\color[HTML]{08090C} Acorus Tatarinowii}                                                      & \begin{tabular}[c]{@{}l@{}}p - Cymene, \\ Nerol acetate\end{tabular}    & \begin{tabular}[c]{@{}l@{}}Open orifices and resolve phlegm, \\ remove dampness \\ and promote appetite\end{tabular}                  \\ \hline
            \cellcolor[HTML]{F2F2F2}{\color[HTML]{08090C} Chinese Angelica}                                                        & Ligustilide                                                             & \begin{tabular}[c]{@{}l@{}}Nourish and activate blood, \\ regulate mood\end{tabular}                                                  \\ \hline
            \cellcolor[HTML]{F2F2F2}{\color[HTML]{08090C} \begin{tabular}[c]{@{}l@{}}Aromatic Turmeric \\ Root-tuber\end{tabular}} & \begin{tabular}[c]{@{}l@{}}Camphene, Camphor,\\  Curcumene\end{tabular} & \begin{tabular}[c]{@{}l@{}}Promote blood circulation \\ to relieve pain, promote \textit{qi} circulation\\ and relieve depression\end{tabular} \\ \hline
            \cellcolor[HTML]{F2F2F2}{\color[HTML]{08090C} Spine Date Seed}                                                         & Suanzaoren fatty acids                                                  & \begin{tabular}[c]{@{}l@{}}Nourish \textit{qi} and \\ tonify the liver, calm the heart \\ and soothe the nerves\end{tabular}                   \\ \hline
            \cellcolor[HTML]{F2F2F2}{\color[HTML]{08090C} Tangerine Peel}                                                          & Hesperidin                                                              & \begin{tabular}[c]{@{}l@{}}Regulate \textit{qi} and \\ nourish the spleen, \\ dry dampness and resolve phlegm\end{tabular}                     \\ \hline
            \end{tabular}
        }
        \vspace{0.1in}
        \footnotetext{(a) Selected TCM}
    \end{minipage}
    \begin{minipage}[]{0.5\textwidth}
        \centering
        \includegraphics[width=1\linewidth]{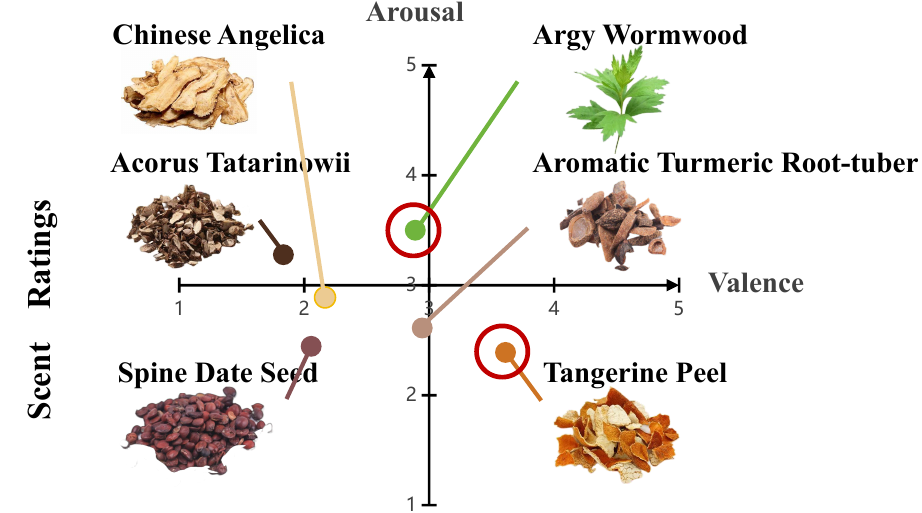}
        \footnotetext{(b) Arousal/valence circumplex}
    \end{minipage}
    \caption{Material and results of Study 1:(a) shows the ingredients and efficacy of the six TCM scents selected in Study 1. (b) presents the results of Study 1, mapping the TCM scents onto the arousal/valence circumplex.}
    \label{Fig.2}
\end{figure}

In this study, we conducted a screening of existing TCM essential oils and selected six that are specifically associated with emotional regulation for further research. Fig.\ref{Fig.2}(a) depicts the six TCM scents, along with their constituent ingredients and efficacy, as outlined in the Pharmacopoeia of the People's Republic of China \cite{2020Pharmacopoeia}.

\subsubsection{Design}
We conducted a within-participants study asking the participants to rate six scents (Argy Wormwood, acorus tatarinowii, Chinese angelica, aromatic turmeric root-tuber, spine date seed, and Tangerine Peel).

\subsubsection{Setup}
Participants sniffed the scents from identical bottles located on a table. Each bottle contained 10ml of essential oil. They sniffed each bottle for 2s, with intervals of 20s(as in \cite{wilson2017multi}).

\subsubsection{Procedure}
After sniffing each scent, participants were asked to rate the valence and arousal of each scent using Self-Assessment Manikins (SAM) \cite{bradley1994measuring} on a 5-point Likert scale (1 = "low", 5 = "high", see Fig.~\ref{Fig.2}(b)).

\subsubsection{Results}
18 participants, 23-26 years old (M = 24.5, SD = 1.04, 8 females), volunteered for this study. 
The normality test showed that the scent ratings were normally distributed. We conducted a repeated measures ANOVA test to analyze the data and found that scents had a main effect on both arousal (F = 2.509, p < 0.05) and valence (F = 9.823, p < 0.001). The results of this study (see Fig. ~\ref{Fig.2}(b)) confirmed that among the six TCM scents, Argy Wormwood (Arousal: M = 3.5, SD = 1.04; Valence: M = 2.89, SD = 0.76) and acorus tatarinowii (Arousal: M = 3.28, SD = 1.45; Valence: M = 1.83, SD = 0.92) had low valence and high arousal. Chinese angelica (Arousal: M = 2.89, SD = 1.18; Valence: M = 2.17, SD = 0.71), aromatic turmeric root-tuber (Arousal: M = 2.61, SD = 1.29; Valence: M = 2.84, SD = 0.99), and spine date seed (Arousal: M = 2.44, SD = 1.2; Valence: M = 2.06, SD = 0.99) had low valence and low arousal. Tangerine Peel (Arousal: M = 2.39, SD = 1.09; Valence: M = 3.61, SD = 1.04) had low arousal and high valence. Since arousal is an important indicator of driving  \cite{jiang2024scented}, we chose Argy Wormwood which has the highest arousal values for Study 2. Considering its valence value is low, we chose Tangerine Peel which has a relatively high valence for comparison.

\subsection{Study 2:Exploring the effects of Argy Wormwood and Tangerine Peel on driving fatigue}

\subsubsection{Design}
This study followed a within-subjects experimental design of 1 (state: fatigued) × 3 (scents: Tangerine Peel, Argy Wormwood, water). It mainly measures the impact of TCM scents on three indicators: alertness, driving performance, and acceptance.

\begin{figure}
\centering
\begin{tabular}{cc}
\includegraphics[width=0.48\textwidth]{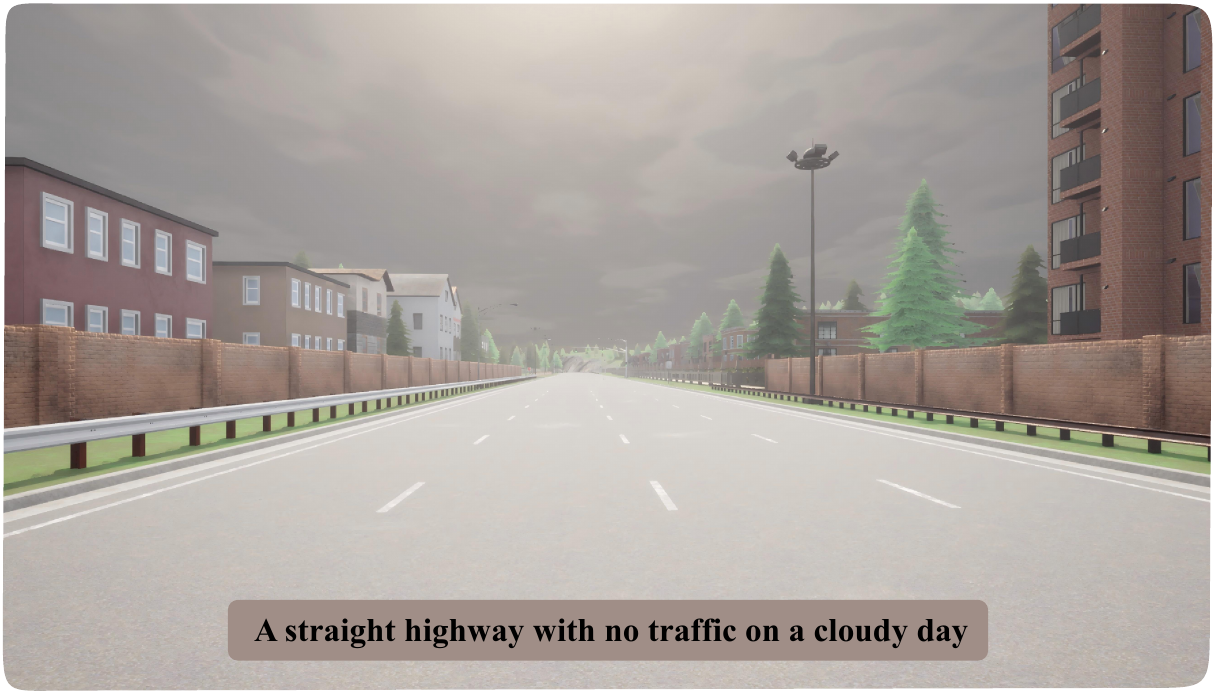} & \includegraphics[width=0.48\textwidth]{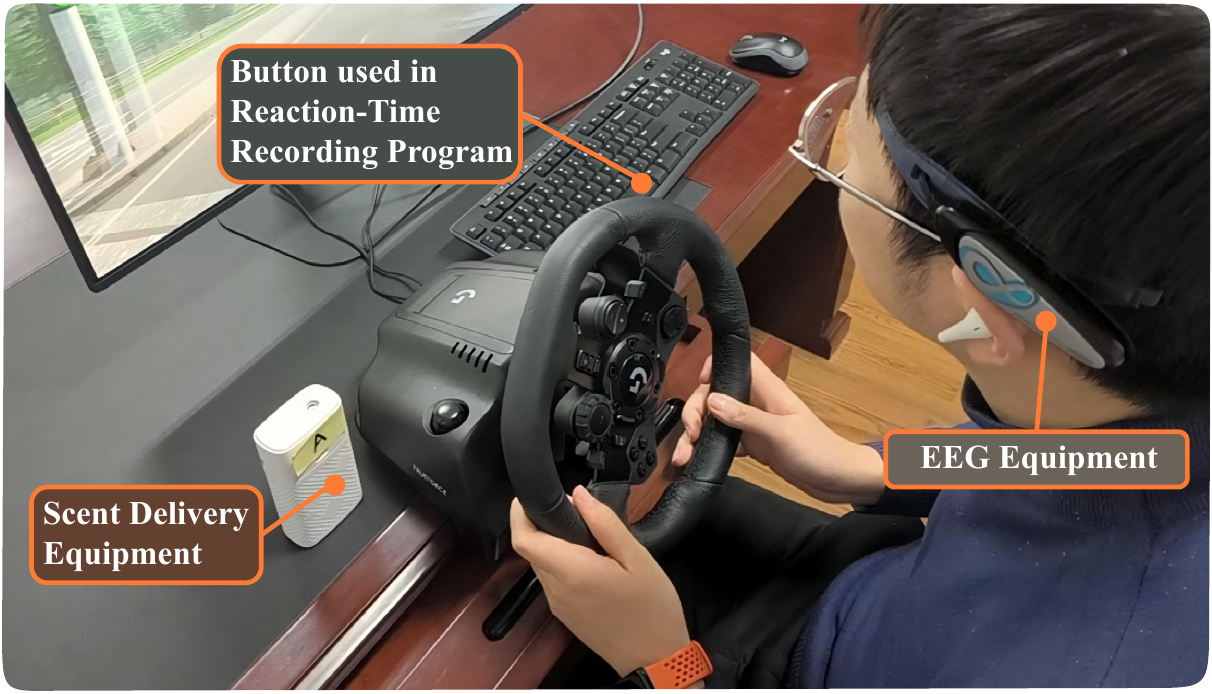} \\
\multicolumn{1}{c}{\small (a) The scenes of the driving simulator} & \multicolumn{1}{c}{\small (b) The location of the experimental equipment}
\end{tabular}
\caption{\small The on-site equipment for Study 2: Fig. (a) shows the driving simulation scene used throughout Study 2. Fig. (b) shows the position of the steering wheel, the EEG equipment, the buttons of the reaction-time recording program, and the scent release device.}
\label{fig:2}
\end{figure}

\begin{itemize}
    \item \textbf{Alertness:} \textit{Karolinska Sleepiness Scale (KSS) \cite{lee2020making}:} A subjective assessment scale for measuring sleepiness levels, with scores ranging from 1 (awake) to 9 (sleepy). The specific design of the KSS scale is detailed in Appendix C. \textit{EEG indicator:} $R(\alpha/\beta)$ was selected as the EEG measurement indicator \cite{s24082384}, a lower ratio indicates higher alertness. The selection of EEG indicators is detailed in Appendix B. 
    \item \textbf{Driving Performance:} \textit{Reaction Time:} Participants were instructed to press a button immediately upon hearing a beep, and their reaction times were recorded.
    \item \textbf{Acceptance:} \textit{Questionnaire:} The questionnaire assessed three dimensions of scent acceptance—usefulness, liking, and comfort \cite{dmitrenko2020caroma}—using a 1–5 scale (1 = low, 5 = high). \textit{Interview:} Interviews explored the impact of TCM scents on driving and participants' willingness to use TCM scents as long-term in-vehicle aromatherapy or short-term reminder scents.
\end{itemize}

\subsubsection{Setup}
\begin{itemize}
    \item \textbf{Driving Simulator:} The driving simulation scenario was constructed using Carla \cite{dosovitskiy2017carla}, featuring a straight, traffic-free highway under cloudy weather conditions (see Fig.~\ref{fig:2}(a)) \cite{huang2024chatbot}. Participants sat in the driving simulator, which was equipped with a Logitech G29 Driving Force steering wheel and pedals (see Fig.~\ref{fig:2}(b)). A 27-inch front main screen with a 60Hz refresh rate displayed the external vehicle view. A Python program was used to record reaction times.
    \item \textbf{Scent Delivery Equipment:} A Nimin aroma diffuser—compatible with the Mi Home APP for adjusting scent emission concentration and interval—was used. Three diffusers contained diluted essential oils (essential oil : alcohol : water = 1 : 6 : 3) of Tangerine Peel, Argy Wormwood, and water, respectively (as in \cite{jiang2023study}). The scent delivery equipment was placed on the left side of the steering wheel.
    \item \textbf{EEG Equipment:} The Brainlink Pro kit, developed by Hongzhili Group, includes a built-in algorithm for calculating focus and relaxation levels, as well as measuring the proportions of alpha, beta, and gamma waves. Pre-experiments revealed that the alpha/beta ratio exhibited the most significant changes during the study; thus, this index was selected as the primary indicator for measuring participants' fatigue levels.
\end{itemize}

\subsubsection{Procedure}
\begin{figure}
    \centering
    \includegraphics[width=1\linewidth]{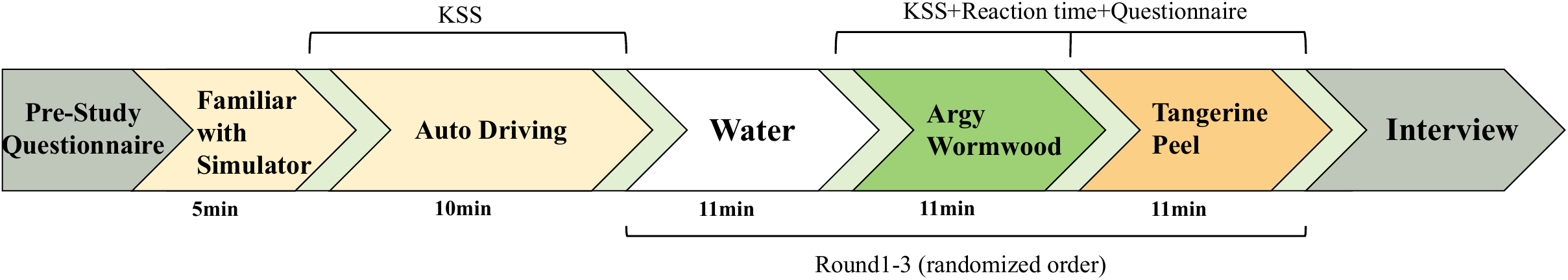}
    \caption{The timeline of the Study 2 procedure. Participants filled out a pre-questionnaire. Then, they drove autonomously for 5min to get used to the simulator. Next, they drove autonomously for 10min to induce fatigue. Then, 3 rounds of scent release experiments were conducted. The round order was randomized. Finally, participants filled out a scent-evaluation questionnaire and had an interview.}
    \label{fig:3}
\end{figure}

As shown in Fig.~\ref{fig:3}, participants first read and signed the informed consent form, then completed a pre-experiment questionnaire to provide basic information such as gender and age. Next, they wore headphones and EEG equipment, tested the headphone volume and reaction-time recording program, and then conducted a 5-minute autonomous driving practice in the simulated highway scenario—with hands on the steering wheel and feet on the brake—to familiarize themselves with the program and driving simulator operations. Following this practice, participants rated their fatigue level using the KSS questionnaire. They then underwent a 10-minute driving fatigue induction experiment, after which they completed the KSS again to verify the induction effect.  

Subsequently, three rounds of scent release experiments were conducted. In each round, scents were released for 30 seconds at 0, 3.5, 7, and 10.5 minutes. At the 11th minute, participants pressed a button upon hearing a beep (with their reaction time recorded), and then filled out the KSS. The order of the three rounds was randomized using a Latin square design to mitigate sequence effects \cite{fisher1930statistical}. Finally, participants completed a scent evaluation questionnaire, participated in a brief interview, and received 45 RMB as compensation for their participation. The total duration of the experiment was approximately 60 minutes.

\subsubsection{Results}
\paragraph{Participants}
50 participants, 19-28 years of age (M = 24.7, SD = 3.64, 25 females).They were asked to be well-rested and avoid caffeine or alcohol before the study.

\begin{figure}[!t]
\centering
\subfloat[KSS before and after driving fatigue induction experiment]{
		\includegraphics[scale=0.45]{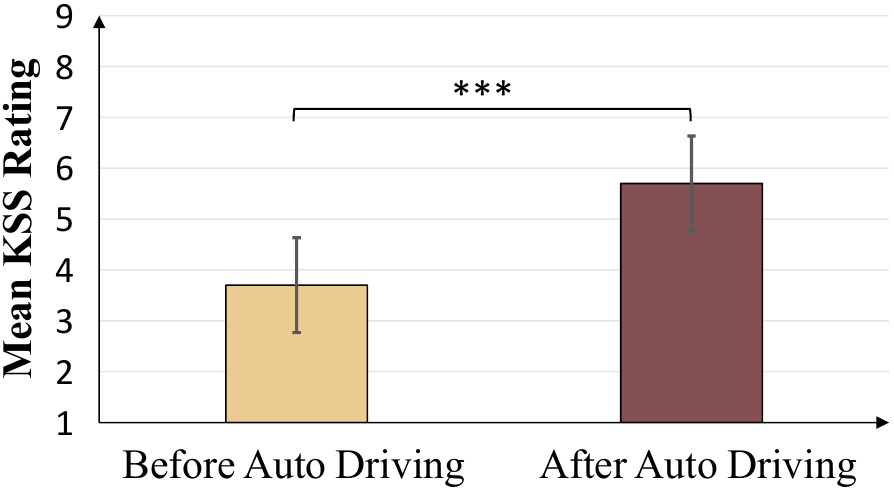}}
\subfloat[KSS after scent release experiments]{
		\includegraphics[scale=0.45]{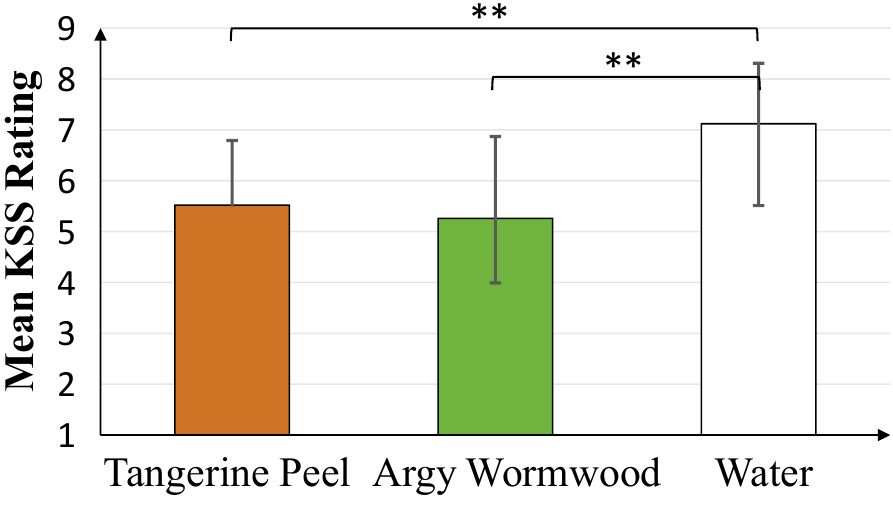}}
\\
\subfloat[EEG ratings after scent release experiments]{
		\includegraphics[scale=0.45]{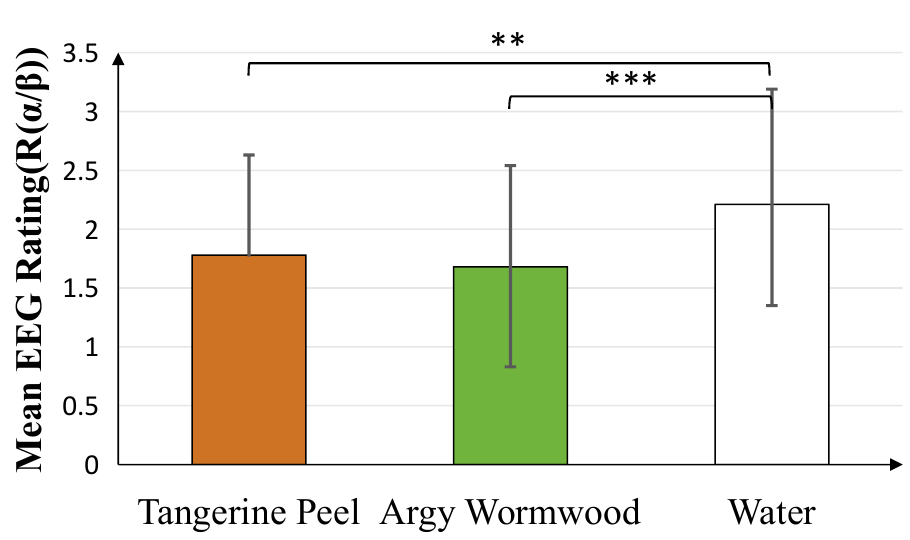}}
\subfloat[Reaction-time after scent release experiments]{
		\includegraphics[scale=0.45]{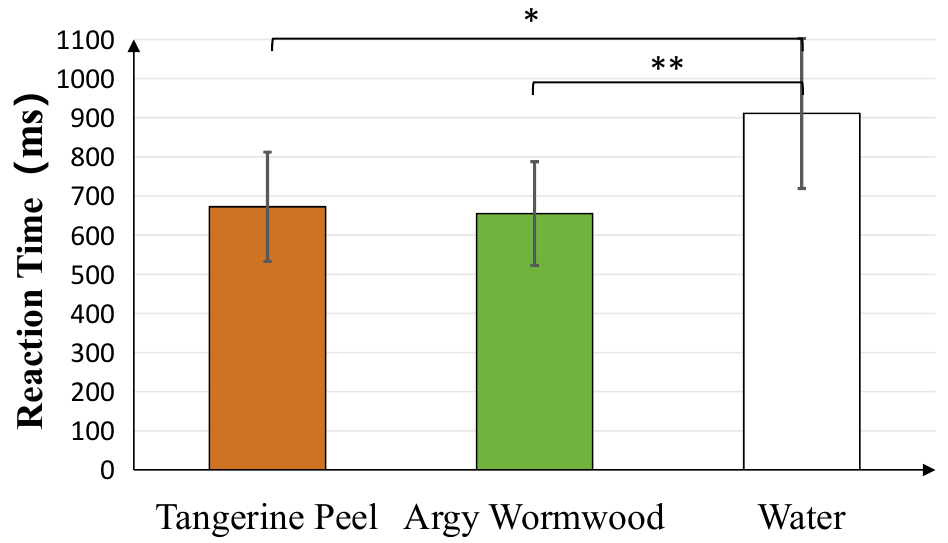}}
\\
\subfloat[Acceptance of TCM scents]{
		\includegraphics[scale=0.43]{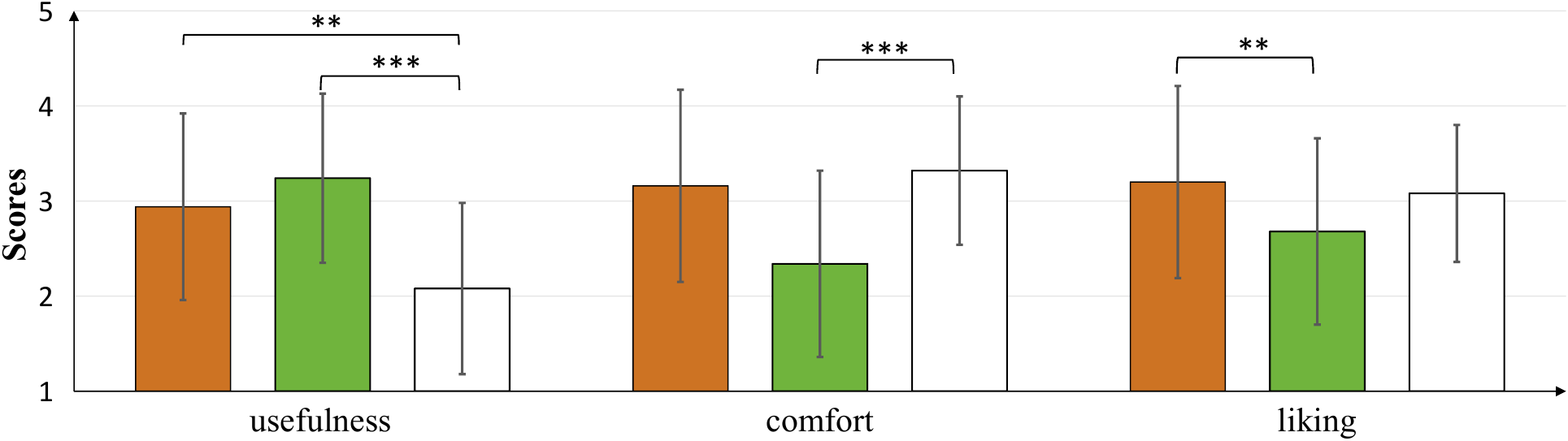}}

\caption{Results of Study 2. (* p < 0.05, ** p < 0.01, *** p < 0.001, all figures follow this annotation)}
\label{fig:4}
\end{figure}

\paragraph{Alertness before and after the driving fatigue induction experiment}
The Shapiro-Wilk test indicated that none of the variables followed a normal distribution. Therefore, the Wilcoxon signed-rank test was employed to compare the mean values between the two variables. Specifically, KSS scores before and after the driving fatigue induction experiment were compared (see Fig.~\ref{fig:4}(a)). The results showed that the KSS score before the experiment (M = 3.7, SD = 0.93) was significantly lower than that after the experiment (M = 5.7, SD = 0.94, Z = 5.13, p < 0.001). This finding suggests that participants were alert at the beginning of the study and reported feeling fatigued following the driving fatigue induction experiment.

\paragraph{Alertness after scent release experiments}
A repeated-measures ANOVA on KSS scores revealed a main effect of scent type on both KSS ratings (F = 3.917, p < 0.05) (see Fig.~\ref{fig:4}(b)) and EEG indices (see Fig.~\ref{fig:4}(c)) (p < 0.001), indicating that scent type can influence drivers' subjective alertness scores. The Friedman test showed that drivers' subjective alertness under the Argy Wormwood and Tangerine Peel conditions was significantly higher than that under the baseline (water) condition (p < 0.05), while no significant difference was observed between the Argy Wormwood and Tangerine Peel conditions (p = 0.29). These results confirm that both TCM scents can improve drivers' subjective alertness. Although the refreshing effect of Argy Wormwood (M = 5.3, SD = 1.61) was slightly stronger than that of Tangerine Peel (M = 5.5, SD = 1.27), the difference was not statistically significant.

\paragraph{Driving Performance after Scent Release Experiments}
A repeated-measures ANOVA on reaction times revealed a significant main effect of scent type on driving performance (F = 3.95, p < 0.05) (see Fig.~\ref{fig:4}(d)), indicating that scent type can notably influence drivers' driving performance. The Friedman test further showed that driving performance under the Argy Wormwood and Tangerine Peel conditions was significantly better than that under the baseline (water) condition (p < 0.05), while no statistically significant difference was observed between the Argy Wormwood and Tangerine Peel conditions (p = 0.85). These results collectively suggest that both TCM scents can effectively improve driving performance, with no statistically significant difference between the two scents.

\paragraph{Acceptance of TCM Scents}
A repeated-measures ANOVA on questionnaire data showed a main effect of scent type on perceived usefulness (F = 8.99, p < 0.01), comfort, and liking, indicating that scent type can affect drivers' acceptance of the scents. Paired-sample tests revealed that the perceived usefulness of Argy Wormwood was significantly higher than that of Tangerine Peel (p < 0.01) (see Fig.~\ref{fig:4}(e)). Among the two TCM scents, Tangerine Peel scored higher than Argy Wormwood in both comfort and liking.  

\begin{figure}[t]
    \centering
    \includegraphics[width=1\linewidth]{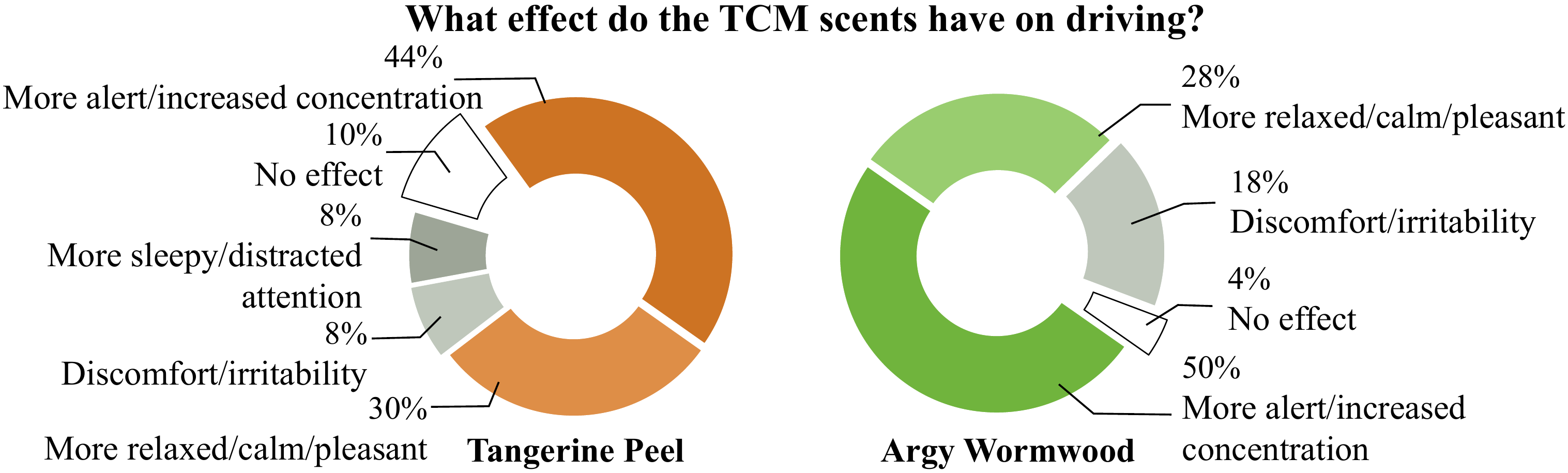}
    \caption{Interview results in Study 2: Most subjects recognized the positive effects of traditional Chinese medicine scents during driving, such as refreshing the mind, improving concentration, or enhancing mood. Among them, Argy Wormwood had a better refreshing effect, while Tangerine Peel was more effective in boosting pleasantness. Regarding the negative effects, a small number of subjects thought that Tangerine Peel could make people sleepier or had no effect, and Argy Wormwood could cause discomfort.}
    \label{fig:5}
\end{figure}

\begin{figure}[t]
    \centering
    \includegraphics[width=0.95\linewidth]{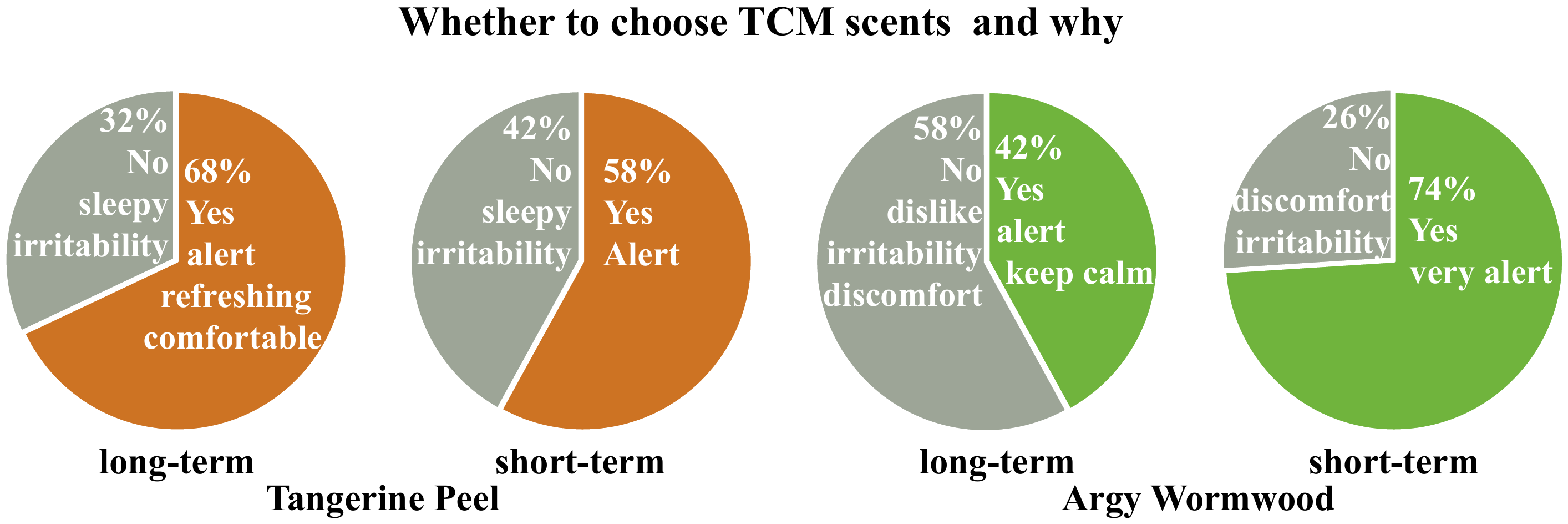}
    \caption{Interview results in Study 2: Most subjects are willing to choose the scent of traditional Chinese medicine as the in-vehicle scent. Due to its certain alerting effect and fresh characteristics, Tangerine Peel is preferred by more subjects as a long-term in-vehicle aromatherapy. Argy Wormwood, with its excellent alerting effect, is favored by most subjects as a short-term reminder scent. However, because of its unpleasant smell, few choose it as a long-term in-vehicle aromatherapy.}
    \label{fig:6}
\end{figure}

Interview results showed that 74\% and 78\% of participants believed that Tangerine Peel and Argy Wormwood, respectively, had positive effects on driving (see Fig.~\ref{fig:5}). Specifically, more participants reported that Argy Wormwood could improve attention compared to Tangerine Peel, while more participants found Tangerine Peel to be pleasant than Argy Wormwood. When asked whether they would choose these two scents as long-term in-vehicle aromatherapy or short-term reminder scents (see Fig.~\ref{fig:6}), most participants preferred Tangerine Peel as a long-term option due to its pleasantness and comfort, and Argy Wormwood as a short-term reminder scent—attributed to its stronger alerting effect, despite being less comfortable.

\section{Discussion and Future Work}

\subsection{Discussion}


\begin{figure}
    \centering
    \includegraphics[width=0.95\linewidth]{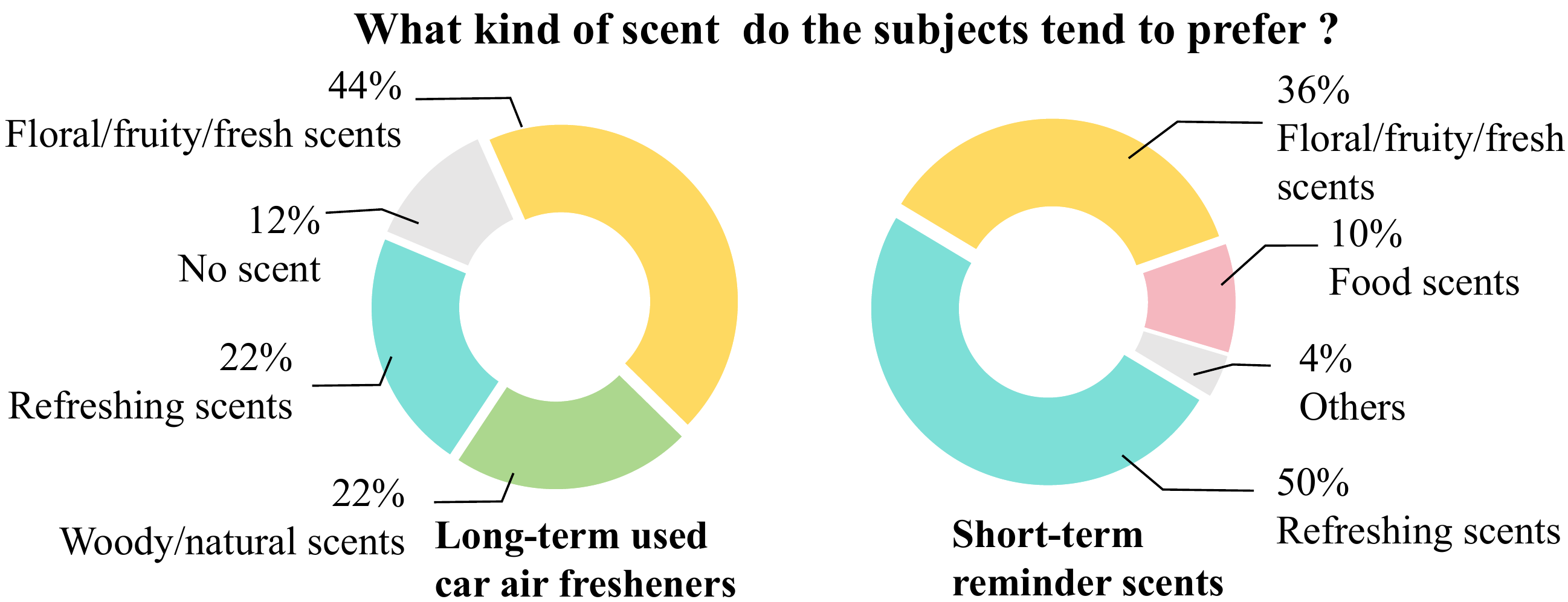}
    \caption{The interview results show that in addition to TCM scents, subjects care more about whether the scent is pleasant when choosing car air fresheners, and pay more attention to the degree of stimulation when selecting reminder scents. Freshness is a scent characteristic mentioned by most subjects.}
    \label{fig:7}
\end{figure}

\subsubsection{Effect of TCM Scents on Driving Fatigue}
The two selected TCM scents both exhibit clear effects in mitigating driving fatigue. They can effectively enhance drivers' alertness, reduce fatigue induced by autonomous driving, and thereby improve driving safety. Between the two, Argy Wormwood demonstrates a slightly stronger effect than Tangerine Peel, though the difference is not statistically significant. This may be attributed to the narrow gap in arousal levels between Argy Wormwood and Tangerine Peel. Additionally, Tangerine Peel, with its higher valence, can provide a more pleasant driving experience.

\subsubsection{Effect of TCM Scents on Driving Performance}
Exposure to the TCM scents resulted in a significant reduction in participants' reaction times, indicating improved driving performance. This outcome is beneficial for enhancing driving safety. Consistent with the findings related to alertness, no statistically significant difference was observed between the effects of Argy Wormwood and Tangerine Peel.

\subsubsection{Acceptance of TCM Scents}
The acceptance levels of the two TCM scents differ significantly. Tangerine Peel exhibits higher ratings in liking and comfort but lower perceived usefulness, while Argy Wormwood shows lower ratings in liking and comfort but higher perceived usefulness. Based on interview results, Tangerine Peel is more suitable for long-term application as an in-vehicle fragrance due to its comforting and subtly refreshing attributes. In contrast, Argy Wormwood is more appropriate as an alertness-enhancing scent during periods of fatigue. It can be activated either actively or passively when drivers experience severe drowsiness, as it possesses a notable refreshing effect despite being less comfortable.

\subsubsection{Limitations}
Due to site and environmental constraints, only Chinese participants aged 20‑30 were recruited. Young drivers were targeted for their greater propensity to adopt the emerging autonomous driving mode. Further investigation is required to clarify whether TCM scents yield similar or distinct effects for drivers of other ages and cultural backgrounds. Moreover, this laboratory simulation differs from real‑world driving contexts. Finally, the selected TCM scents were determined by their efficacy and the arousal‑valence model, leaving numerous other TCM types unexamined for future study.

\subsection{Future Work}
\subsubsection{What Characteristics of Scents Make Them More Suitable for Long-Term or Short-Term In-Vehicle Aromatherapy?}
According to post-experiment open interviews (see Fig.~\ref{fig:7}), most participants indicated a preference for choosing scents they liked and that had fresh attributes—such as floral, fruity, and woody scents—as long-term in-vehicle aromas. For short-term in-vehicle aromas, participants prioritized the refreshing effect (i.e., irritancy) of scents. They typically selected scents with strong refreshing properties, such as essential balm, floral water, and cooling ointment, or pungent food odors like durian and snail noodles. Building on this insight, future research could explore the arousal and valence levels of more TCM scents, classifying those with high arousal as short-term in-vehicle aromas and those with high valence as long-term options. This would expand the range of scents available for driving fatigue regulation and enable more refined, personalized design.

\subsubsection{What Factors Influence the Refreshing Effect of Scents?}
To validate the experimental findings and inform future studies, we conducted post-experiment interviews on scent familiarity and perceived characteristics. Most participants described Tangerine Peel as sweet-sour and fruity, and Argy Wormwood as woody and reminiscent of TCM. Approximately 25\% correctly identified Argy Wormwood, mainly due to prior cultural or everyday exposure. Whether such familiarity influences the refreshing effect warrants further investigation.

Participants also reported that alertness was strongest upon the first scent release and diminished with repeated exposure to the same scent, suggesting that the refreshing effect may partly arise from scent transitions. Alternating between different scents may therefore be more effective than repeatedly releasing a single scent.

\subsubsection{AI-Integrated TCM Scent System for Autonomous Driving Fatigue Mitigation}
Artificial intelligence (AI) can determine whether a driver is fatigued and the severity of fatigue by analyzing in-vehicle and external environmental data. It can use dashboard camera data to assess if the driver is engaged in prolonged monotonous driving, and in-vehicle cameras to recognize facial expressions \cite{zhu2022research} or upper-body posture changes \cite{ansari2022automatic} as indicators of fatigue. Based on the driver’s fatigue state and the characteristics of different TCM scents, AI can generate tailored adjustment strategies to mitigate driving fatigue.

\section{Conclusion}
This paper demonstrates that Traditional Chinese Medicine (TCM) scents can effectively mitigate fatigue induced by autonomous driving, while also enhancing drivers' alertness and driving performance. Specifically, Argy Wormwood exhibits a stronger refreshing effect but receives lower ratings in terms of liking and comfort, making it more suitable as a short-term alerting scent. In contrast, Tangerine Peel has a relatively milder refreshing effect but achieves higher scores in liking and comfort, rendering it more appropriate for long-term use as an in-vehicle aroma. As a pioneering exploration into the application of TCM scents for driving fatigue mitigation, our study provides valuable insights for the future development of in-vehicle comfort and safety systems.

\bibliographystyle{ACM-Reference-Format}
\bibliography{ref}

\appendix
\newpage
\begin{table}[!htbp]
\centering
\caption{The overview of appendix} 
\label{tab:tableTab} 
\begin{tabular}{cc} 
\toprule 
appendix   & page  \\
\midrule 
Participant details & 14  \\
Pre-study & 15  \\
Scale Design & 17 \\
Interview Design and Supplementary Conclusions & 18 \\
\bottomrule 
\end{tabular}
\end{table}

\section{Participant details}
\begin{figure}[h]
    \centering
    \includegraphics[width=0.76\linewidth]{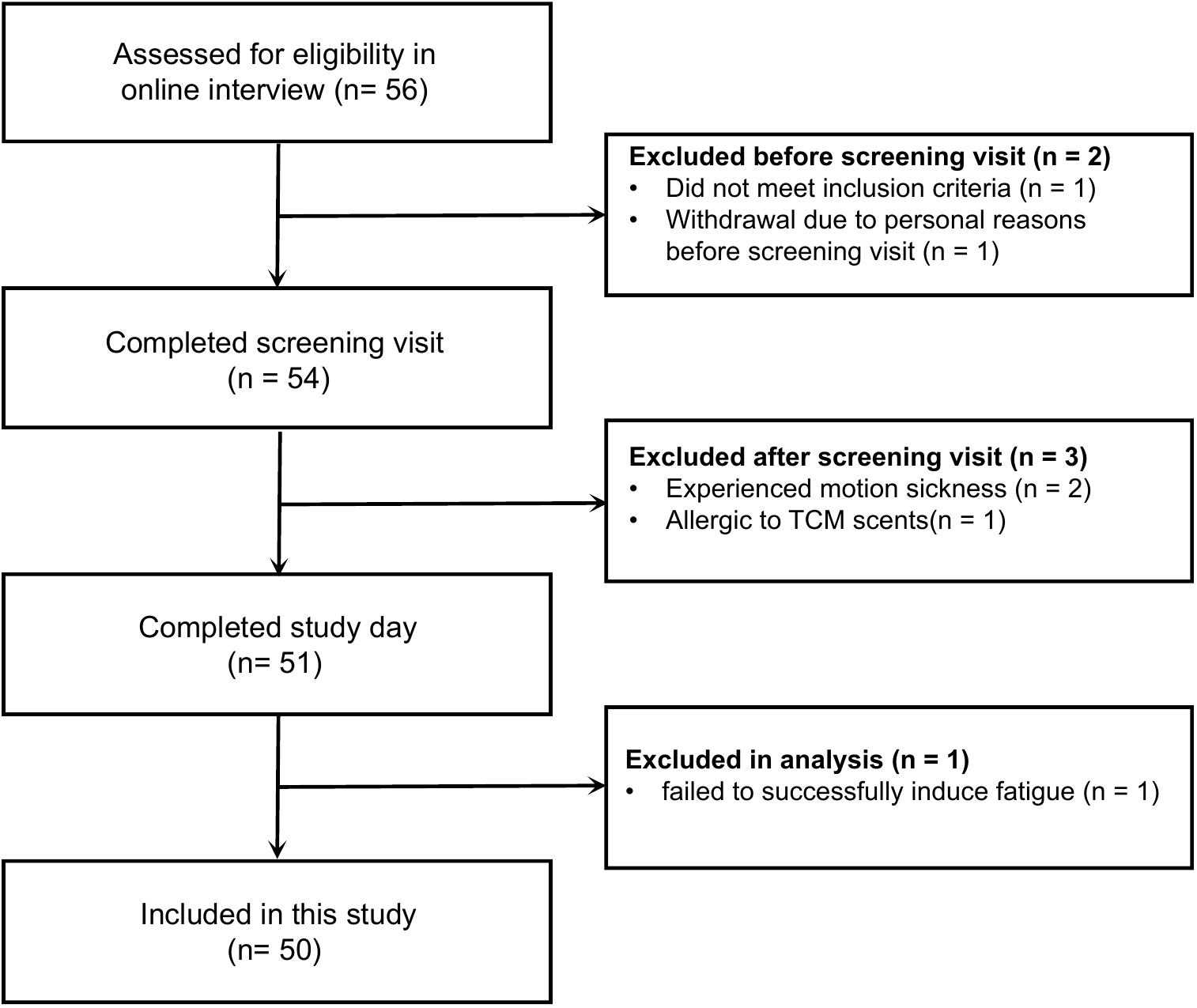}
    \caption{ Study flow diagram. Participants subject to screening and inclusion/exclusion in study 2.}
    \label{fig:app-3}
\end{figure}

The study flow diagram is shown in  Fig.~\ref{fig:app-3}. Inclusion and exclusion criteria for participation are explained in the following.

\paragraph{Inclusion criteria for individuals eligible for participation in study 2}
(1)aged between 20 and 30 years old;(2)possession of a driver’s license that is valid in China.

\paragraph{Exclusion criteria for participation in study 2}
 Having respiratory diseases such as rhinitis, and being allergic to the scents of TCM; having ingested foods that may affect nervous system function, alertness levels, and emotional states (e.g. caffeine, alcohol) within 24 hours before the study; being unable to drive using the driving simulator, or experiencing motion sickness (e.g., dizziness, nausea, etc.) during the simulator test drive session; being unable to follow the study procedures, for example, due to language barriers, mental illnesses, dementia, etc; failure to induce fatigue.

\newpage

\section{Pre-study}
In order to explore the proper duration of the driving fatigue induction experiment, the scent concentration of the scent release experiment, and the EEG indicator in Study 2, we designed two Pre-studies to conduct the exploration. 

\subsection{Pre-study 1: exploring the proper duration of the driving fatigue induction experiment and the scent concentration of the scent release experiment}
\subsubsection{Design}
The experimental procedure is the same as that of Study 2. The questionnaire only collects basic personal information, and no interviews are conducted.

\subsubsection{Setup}
The experimental equipment and the on-site settings are kept consistent with those in Study 2.

\subsubsection{Procedure}
\begin{figure}[h]
    \centering
    \includegraphics[width=1\linewidth]{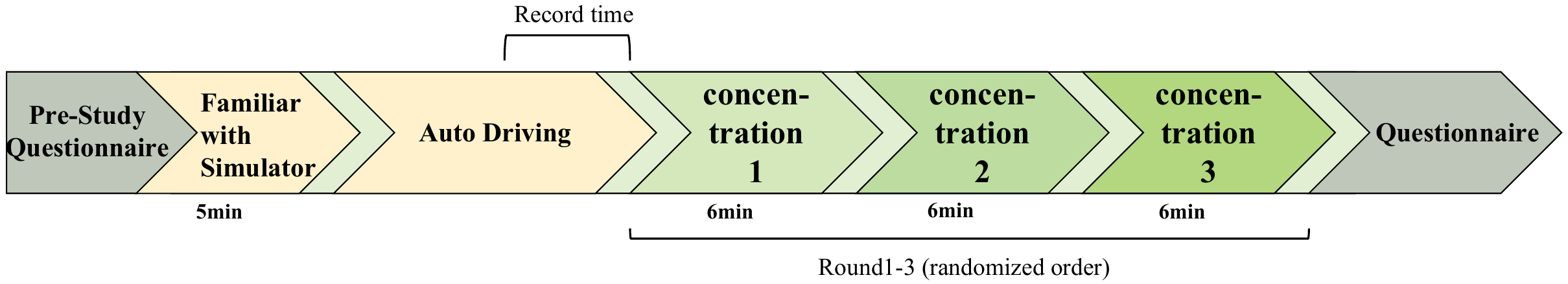}
    \caption{The timeline of the pre-study procedure. Participants filled out a pre-questionnaire. Then, they drove autonomously for 5min to get used to the simulator. Next, they drove autonomously and recorded the time when they were induced to driving fatigue. Then, 3 rounds of scent release experiments were conducted. The round order was randomized. Finally, participants filled out a questionnaire to choose the most comfortable round.}
    \label{fig:app-1}
\end{figure}

As in Fig.~\ref {fig:app-1}, the participants read, signed the consent form, and completed a pre-questionnaire on basic information such as gender and age. Then, they drove autonomously for 5min in a simulated highway scene, hands on the wheel and feet on the brake to get familiar with the driving simulator requirements. Next, they engaged in a driving fatigue induction experiment and recorded the time when they reached a state of general drowsiness and extreme drowsiness. The simulator scenario was the same as that in Study 2. The maximum duration of this experiment is 25min. If the participants still have not reached a state of extreme fatigue, the experiment will be stopped, and the data will be marked as "failure to induce fatigue".  After that, three rounds of scent release experiments were administered. The concentration of scent in each round was controlled by regulating the scent release time each time. The release times for the three rounds were 8s every three minutes, 16s every three minutes, and 30s every three minutes respectively. The round order was randomized by Latin-square design to mitigate sequence

\subsubsection{Results}
\paragraph{Participants}7 participants, 23-25 years old (M = 24.1, SD = 1.21, 3 females), volunteered for this study.

\paragraph{The proper duration of the driving fatigue induction experiment}Since the data of one participant was marked as "failure to induce fatigue", we only calculated the mean and variance of six groups of data. The calculation shows that the average time for participants to reach a state of general fatigue is 3.5min (SD = 1.87), and the average time to reach a state of extreme fatigue is 7.5min (SD = 1.38). In order to ensure that the participants are induced to experience driving fatigue, we selected 10min as the duration of the driving fatigue induction experiment in Study 2.

\paragraph{The proper concentration of the scent release experiment}Among the seven experiment participants, two chose the lowest concentration because they were allergic to the smell of traditional Chinese medicine, and a high concentration might irritate them and cause sneezing. The remaining five participants chose the highest concentration because of the short duration of scent effectiveness. Therefore, in the formal experiment, we screened out the participants who were allergic to the TCM scents and selected the highest concentration as the experimental concentration for Study 2. Due to the limitations of the equipment, we were unable to explore the impact of higher concentrations on the participants. This may be further studied in future experiments.

\subsection{Pre-study 2: Exploring the proper EEG indicator in Study 2}
Common EEG indicators for detecting fatigue include the R (alpha/beta) . The BrainLink Pro accessory selected in the experiment is equipped with a built-in algorithm to calculate "attention". A pre-study was conducted to test whether these two indicators have the effect of indicating fatigue.

\subsubsection{Design}
The experimental design is the same as that of Study 2. The questionnaire only collects basic personal information, and no interviews are conducted.

\subsubsection{Setup}
The experimental equipment and the on-site settings are kept consistent with those in Study 2.

\subsubsection{Procedure}
\begin{figure}[h]
    \centering
    \includegraphics[width=1\linewidth]{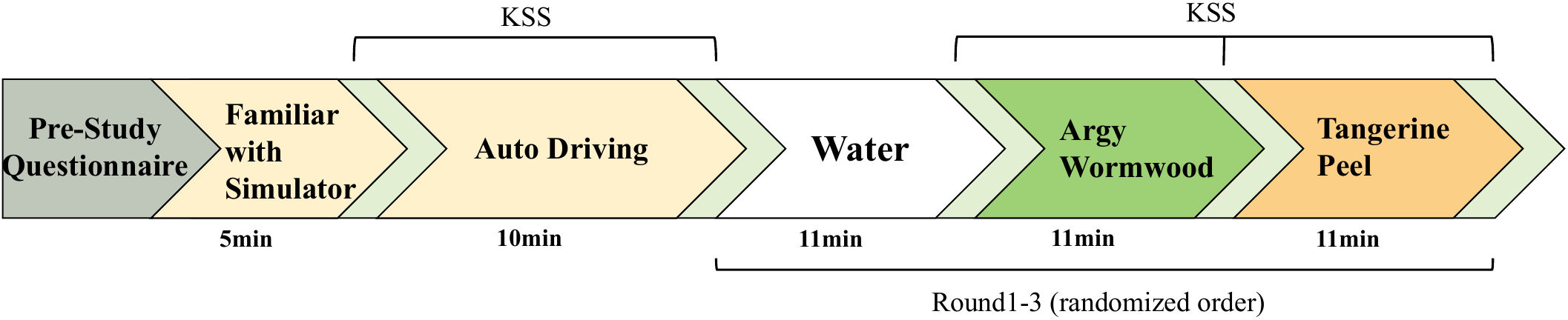}
    \caption{The timeline of the pre-study 2 procedure. Participants filled out a pre-questionnaire. Then, they drove autonomously for 5min to get used to the simulator.  Then, 3 rounds of scent release experiments were conducted. The round order was randomized.}
    \label{fig:app-2}
\end{figure}

As in Fig. ~\ref{fig:app-2}, participants read, signed the consent form, and filled out a pre-questionnaire on basic info like gender and age. Then, they put on headphones and EEG equipment and drove autonomously for 5min in a simulated highway scene, hands on the wheel and feet on the brake to get familiar with the program and driving simulator requirements. Next, they used the KSS questionnaire to rate fatigue. Then, participants engaged in driving fatigue induction experiment for 10min. After that, they filled out the KSS again to confirm the effect. Subsequently, three rounds of scent release experiments were administered. Scents were released at 0, 3.5, 7, and 10.5min for 30s in each round. The round order was randomized by Latin-square design to mitigate sequence effects \cite{fisher1930statistical}. 

\subsubsection{Results}
\paragraph{Participants}6 participants, 22-25 years old (M = 22.7, SD = 1.05, 3 females), volunteered for this study.

\paragraph{The proper EEG indicator of Study 2}The experiment showed that there is significant individual variability in the two values—different participants often have different individual average values. However, when compared uniformly, the R (alpha/beta) ratio has a better indicative effect than the "attention level". Additionally, the fluctuation of this objective indicator is in good consistency with the subjective Karolinska Sleepiness Scale (KSS). Therefore, the R (alpha/beta) ratio was selected as the EEG indicator.

\section{Scale Design}
\subsection{Karolinska Sleepiness Scale}
\begin{table}[!htbp]
\centering
\caption{Scale Level and description of Karolinska Sleepiness Scale} 
\label{tab:tableTab} 
\begin{tabular}{cp{12cm}} 
\toprule 
Scale Level   & Description  \\
\midrule 
1 & Extremely alert   \\
2 & Very alert  \\
3 & Alert \\
4 & Rather alert \\
5 & Neither alert nor sleepy\\
6 & Some signs of sleepiness\\
7 & Sleepy, but no effort to keep awake\\
8 & Sleepy, but some effort to keep awake\\
9 & Very sleepy, great effort to keep awake, fighting sleep \\
\bottomrule 
\end{tabular}
\end{table}

\subsection{Usefulness}
\begin{table}[!htbp]
\centering
\caption{Scale Level and description of Usefulness} 
\label{tab:tableTab} 
\begin{tabular}{cp{12cm}} 
\toprule 
Scale Level   & Description  \\
\midrule 
1 & Feeling more fatigued after smelling the scent   \\
2 & No difference in fatigue level before and after smelling the scent  \\
3 & Slight fatigue-alleviating effect, which is present when the scent is just sprayed but fades after 3-5 seconds \\
4 & Relatively sustained fatigue-alleviating effect, lasting 1-2 minutes per spray \\
5 & Highly sustained fatigue-alleviating effect, with the scent detectable throughout after spraying\\
\bottomrule 
\end{tabular}
\end{table}

\newpage
\section{Interview Design and Supplementary Conclusions}
\begin{table}[!htbp]
\centering
\caption{The overview of interview} 
\label{tab:tableTab} 
\begin{tabular}{cp{10cm}} 
\toprule 
Inquiry Aspects	& Specific Examples  \\
\midrule 
Scent Familiarity &	Have you smelled this scent before? Can you roughly describe its characteristics? Did it trigger any memories for you?\\
Impact of Scent on Driving &	Did the TCM scent have any impact on other emotions or your driving state while driving, such as making you more pleasant or unpleasant, more relaxed or more focused?\\
Willingness to Choose &	Would you choose this scent as a long-term in-car aromatherapy or a short-term reminder scent? Please explain why. If not, what scenarios do you think it is more suitable for?\\
Preferred Scent Characteristics &	What characteristics of scents would you prefer for in-car aromatherapy or short-term reminder scents?\\

\bottomrule 
\end{tabular}
\end{table}
\subsection{Scent Familiarity}
This question was mainly intended to confirm whether subjects could smell the scents accurately, so as to prevent the experimental results from being affected. Among the 50 subjects included in the measurement, 13 accurately identified the scent of Argy Wormwood, while the rest described it as a bitter traditional Chinese medicine or herbal smell; none accurately identified the scent of Tangerine Peel, but 10 described it as a citrus fruit smell, and the rest as a sweet and sour fruit smell. This indicates that all subjects could accurately smell and recognize the characteristics of the scents, and their familiarity with Argy Wormwood was higher than that with Tangerine Peel. This is because Argy Wormwood is a commonly used traditional Chinese medicine in daily life, widely applied in foot therapy, massage, moxibustion and other daily health care. During the traditional Chinese Dragon Boat Festival, there are also customs of placing Argy Wormwood and eating eggs boiled with Argy Wormwood. A small number of subjects stated that smelling Argy Wormwood reminded them of family reunions or elders at home, thus making them feel more calm.
\subsection{Impact of Scent on Driving}
This conclusion has already been elaborated in the main text. In the open-ended questions, some subjects mentioned that the refreshing effect of the first spray was stronger than that of subsequent sprays. This may be because prolonged spraying led to habituation to the scent, thus failing to further highlight the refreshing effect. This indicates that spraying duration and concentration can affect the experimental results, and further experiments can be designed accordingly. In addition, some subjects suggested that the refreshing effect might be caused by scent transition rather than a specific scent itself, and that alternating sprays of multiple scents might yield better results—this still requires verification through further experiments.
\subsection{Willingness to Choose}
This conclusion has been fully elaborated in the main text and will not be repeated here.  

\subsection{Preferred Scent Characteristics}
This question was only asked when subjects did not choose the two TCM scents as long-term or short-term options, resulting in a small sample size for this part of the survey. However, the conclusion in the main text can still be drawn, as the reasons for choosing scents for long-term or short-term use were consistent: long-term scents tended to be fresh and preferred ones, while short-term scents were inclined to be stimulating, such as common mosquito repellents like essential balm and cooling ointment, or smells of pungent foods like durian and stinky tofu. These stimulating scents may have a good reminder effect in a short time, but they are not necessarily suitable for in-vehicle use, as they might trigger other negative emotions or motion sickness. Further experiments are needed to verify this.
\end{document}